\documentclass{svproc}

\usepackage{amsmath}
\usepackage{amssymb}
\usepackage[T1]{fontenc}
\usepackage{graphicx}
\usepackage{subcaption}
\usepackage[utf8]{inputenc}
\usepackage{listings}
\usepackage{todonotes} 

\usepackage{tikz}
\usetikzlibrary{automata, positioning}

\usepackage{xcolor}
\definecolor{darkblue1}{rgb}{0, 0, 1.0}
\definecolor{darkred1}{rgb}{1.0,0.01,0.15}
\definecolor{blue1}{rgb}{0.2, 0.4, 0.8}
\definecolor{blue2}{rgb}{0.4, 0.6, 0.9}
\definecolor{blue3}{rgb}{0.6, 0.7, 1.0}
\definecolor{red1}{rgb}{0.9, 0.2, 0.2}
\definecolor{red2}{rgb}{0.8, 0.4, 0.4}
\definecolor{red3}{rgb}{1.0, 0.6, 0.6}
\definecolor{purple1}{rgb}{0.6, 0.2, 0.8}
\definecolor{purple2}{rgb}{0.7, 0.4, 0.9}
\definecolor{purple3}{rgb}{0.8, 0.6, 1.0}
\definecolor{green1}{rgb}{0.01, 1.0, 0.02}
\definecolor{green2}{rgb}{0.4, 0.9, 0.4}
\definecolor{green3}{rgb}{0.6, 1.0, 0.6}

\usepackage{txfonts}

\newcommand{\Rn}{\mathbb{R}^{N}}

\newcommand{\R}{\mathbb{R}}
\newcommand{\abs}[1]{\left|#1\right|}  
\newcommand{\norm}[1]{\left\lVert#1\right\rVert}

\newcommand{\nats}{\mathbb{N}}
\newcommand{\reals}{\mathbb{R}}
\newcommand{\comps}{\mathbb{C}}
\newcommand{\ovec}[1]{\mathbf{#1}}
\newcommand{\omat}[1]{\mathbf{#1}}
\newcommand{\oalgst}[1]{\left[#1\right]}
\newcommand{\ofunc}[3]{{#1} : {#2} \longrightarrow {#3}}

\newcommand{\scal}{\mathcal{S}}

\title{Consensus in Models for Opinion Dynamics with Generalized-Bias}

\author{Juan Paz\inst{1} \and Camilo Rocha\inst{1} \and Luis Tobón\inst{1} \and Frank Valencia\inst{2,1}
}

\institute{Pontificia Universidad Javeriana Cali, Colombia\\
\and
CNRS-LIX, École Polytechnique de Paris, France.\\
}

\begin{document}

\maketitle

\begin{abstract}
  Interest is growing in social learning models where users share
  opinions and adjust their beliefs in response to others.
%
  This paper introduces generalized-bias opinion models, an extension
  of the DeGroot model, that captures a broader range of cognitive
  biases.
  These models can capture, among others, dynamic (changing) influences as well as in-group favoritism and out-group hostility, a bias
  where agents may react differently to opinions from members of their
  own group compared to those from outside. 
%
  The reactions are formalized as arbitrary functions that depend, not
  only on opinion difference, but also on the particular opinions of the
  individuals interacting.
  Under certain reasonable conditions, all agents --despite their
  biases-- will converge to a consensus if the influence graph is
  strongly connected, as in the original DeGroot model.
  The proposed approach combines different biases, providing deeper
  insights into the mechanics of opinion dynamics and influence within
  social networks.

\end{abstract}

\section{Introduction}
\label{sec.intro}

Social networks have played a significant role in shaping users'
opinions, often influencing democratic processes and contributing to
social polarization.
Broadly, the dynamics of opinion formation in social networks involve
users expressing their views, encountering the opinions of others, and
potentially updating their own beliefs based on these
interactions.
There is a growing interest in developing models of social learning
~\cite{Wasserman1994} capturing these dynamics to gain insights on how
opinions form and spread within social networks.

The DeGroot model~\cite{degroot1974} is perhaps the most prominent
framework for social learning and opinion formation dynamics. In this
model, a community is represented as a weighted directed graph, known
as the \emph{influence graph}, with edges indicating how much
individuals (\emph{agents}) influence one another.
Each agent has an opinion represented as a value in the interval
$[0,1]$, indicating the strength of the agreement with an underlying
proposition (e.g., ``\emph{AI poses a threat to humanity}''). Agents
revise their opinions by ``averaging'' them with those of their
contacts, weighted by their influence. Studies support the validity of
opinion formation, in many cases, by averaging as demonstrated in
controlled sociological experiments~\cite{degrootEmpirico2015}.

A key theoretical result about the model states that the agents will
converge to \emph{consensus} if the influence graph is \emph{strongly
connected} and the agents have non-zero self-influence (\emph{puppet
freedom} is assumed)~\cite{Golub2017survey}. The significance of this
result lies in the fact that consensus is a central problem in social
learning.
Indeed, the inability to reach consensus is a sign of a polarized
community. The DeGroot model is valued for its tractability, derived
from its connection with matrix powers and Markov chains; it remains a
significant focus of study for understanding of opinion
evolution~\cite{Golub2017survey}.

The DeGroot model, however, assumes two constraints that could be
overly demanding in the context of social networks. It assumes
\emph{homogeneity} and \emph{linearity} of opinion update dynamics.
In social scenarios, two agents may update their opinions differently
depending on their individual \emph{cognitive biases} on disagreement
(i.e., on how they interpret and react towards the level of
disagreements with others). This results in more involved updates that
may rely on non-linear, even non-monotonic, functions.  For example,
an individual under \emph{confirmation (cognitive)
bias}~\cite{Aronson10} may ignore the opinion of those whose level of
disagreement is over a certain threshold.
A recent work~\cite{alvim-multiagentopinion-forte2024} introduces an extension of the DeGroot
model that allows for some form of \emph{heterogeneous} and
\emph{non-linear} opinion updates.
The model  takes \emph{disagreement} between agents as
one of its key parameters.

Nevertheless, representing cognitive biases solely as functions of
disagreement~\cite{alvim-multiagentopinion-forte2024} overlooks key behaviors where reactions
to the same opinion can vary greatly depending on whether the opposing
view comes from someone within their own political group.
%
This can be attributed to \emph{in-group favoritism} (or out-group
hostility), which are aspects of inter-group bias and social identity
theory~\cite{tajfel_turner_1979}. In order to capture these nuances,
models of cognitive biases need to incorporate factors beyond mere,
isolated, disagreement.

This paper introduces \emph{generalized-bias opinion models}, a
framework that generalizes the aforementioned models: it can
capture a broader range of cognitive biases formalized as \emph{arbitrary}
functions that depend, not only on opinion disagreement, but also
on the opinions of the individuals involved.
It is demonstrated that the proposed framework can effectively
represent standard cognitive biases of significant relevance in social
networks. These include, among others, confirmation and identity-based
\emph{inter-group biases}~\cite{tajfel_turner_1979}, where users are
more inclined to favor opinions from people within their own
ideological group. It also allows for modelling \emph{dynamic} (changing) influences.
The approach allows for the combination and integration of these
biases, enabling a more comprehensive and formal understanding of
opinion dynamics and influence in social networks.

Furthermore, the key result of the DeGroot model is extended by
demonstrating that, under certain reasonable conditions, all
agents—despite being subject to the aforementioned cognitive
biases—will converge to a consensus if the influence graph is strongly
connected. This highlights the robustness of the proposed framework in
achieving consensus even when accounting for complex, arbitrary bias-driven
dynamics under some reasonable conditions. Additionally, it is also shown that the proposed framework
allows for more expressive models to those introduced
in~\cite{Alvim2023,alvim-multiagentopinion-forte2024,degroot1974}. These results contribute to a
richer representation of human cognition and interaction in social
networks. Indeed, the framework is illustrated with a case-study modelling inter-group bias.

\section{Preliminaries}
\label{sec.prelim}

This paper uses notation and background from linear
algebra~\cite{Meyer2000}. This section summarizes the main aspects of
the notation and the background. Further details on special types of
matrices and dynamical systems~\cite{Bernardo2008} can be found in
Appendix~\ref{sec.prelimapp}.

The term $N$ is a positive integer and is used to identify the number
of agents in a network.
The expressions $\reals^N$ and $\reals^{N\times N}$ denote the
collections, respectively, of real-valued vectors of length $N$ and
real-valued squared matrices of side $N$. Boldface lowercase letters
identify vectors in $\reals^N$ and boldface uppercase letters identify
matrices in $\reals^{N \times N}$.
The expression $\omat{I}$ denotes the identity matrix in $\reals^{N
  \times N}$ and $\mathrm{det}(\_)$ the determinant (partial) function
on matrices.
Recall that if the inverse $\omat{A}^{-1}$ of $\omat{A}$ exists, it
satisfies $\omat{A}\omat{A}^{-1} = \omat{A}^{-1}\omat{A} =
\omat{I}$.
The L1 norm of a vector $\ovec{x}$ is denoted $\norm{\ovec{x}}_1$.
When the values of a vector $\ovec{x}$ need to be made explicit, it
can be denoted as $\ovec{x} = \oalgst{x_i}_{i \in I}$ or
$\oalgst{x_i}$ when the index set $I$ can be inferred from the
context.
Similarly, when the values of a matrix $\omat{A}$ need to be made
explicit, it can be denoted as $\omat{A} = \oalgst{a_{ij}}_{(i,j) \in
  I\times J}$ or $\oalgst{a_{ij}}$ when the index sets $I$ and $J$ can
be inferred from the context.
It is said that $\omat{A}$ is \emph{non-negative}, denoted $\omat{A}
\geq 0$, iff $a_{ij} \geq 0$ for $1 \leq i, j \leq N$. Likewise,
$\omat{A}$ is \emph{positive}, denoted $\omat{A} > 0$, iff $a_{ij} >0$
for $1 \leq i,j \leq N$.
A matrix $\omat{A}$ is primitive iff $\omat{A}^m > 0$ for some $m \in
\nats$~\cite{Meyer2000}.
A (square) matrix $\omat{P}$ is called a \emph{permutation matrix}
when it has exactly one entry of 1 in each row and each column, and 0s
elsewhere.  A matrix $\omat{A}$ is \emph{reducible} if there is a
permutation matrix $\omat{P}$ such that
$\omat{P}^T\omat{A}\omat{P} =
\begin{bmatrix}
  \omat{B} & \omat{C} \\
  \omat{0} & \omat{D}
\end{bmatrix}$
where $\omat{B}$ and $\omat{D}$ are square matrices (possibly of
different sizes), $\omat{C}$ is a matrix of appropriate size, and
$\omat{0}$ is the zero matrix of appropriate size.
If no such permutation matrix $\omat{P}$ exists, then $\omat{A}$ is
called \emph{irreducible}.

A non-zero vector $\ovec{x}$ is an \emph{eigenvector} of a matrix
$\omat{A}$ with \emph{eigenvalue} $\lambda \in \comps$ iff
$\omat{A}\ovec{x} = \lambda\ovec{x}$ (i.e., applying $\omat{A}$ to the
eigenvector $\ovec{x}$ only scales the eigenvector by the scalar value
$\lambda$).
The \emph{spectrum} of $\omat{A}$ is the set $\sigma(\omat{A})$ of
all eigenvalues $\lambda$ satisfying the equation
$\mathrm{det}(\omat{A} - \lambda\omat{I}) = 0$.
The \emph{geometric multiplicity} of $\lambda \in \sigma(\omat{A})$,
denoted $m_g(\lambda)$, is the number of eigenvectors associated to
$\lambda$ in the (spectral) decomposition of $\omat{A}$.
The \emph{algebraic multiplicity} of $\lambda \in \sigma(\omat{A})$,
denoted $m_a(\lambda)$, is the number of times $\lambda$ appears as a
root of the characteristic polynomial of $\omat{A}$.
It is known that $m_g(\lambda) \leq m_a(\lambda)$ for any $\lambda \in
\sigma(\omat{A})$.
The \emph{spectral radius} of $\omat{A}$ is the real value
$\rho(\omat{A})$ defined by $\rho(\omat{A}) = \max_{\lambda \in
  \sigma(A)} \abs{\lambda}$ identifying the greatest magnitude among
all eigenvalues of $\omat{A}$. If $\lambda \in \sigma(\omat{A})$
and it is the only eigenvalue that satisfies $|\lambda| =
\rho(\omat{A})$, then $\lambda$ is called the \emph{dominant
eigenvalue} of $\omat{A}$.

The \emph{Perron-Frobenius Theorem}~\cite{Meyer2000} is a fundamental
result about non-negative matrices. For irreducible $\omat{A} \in
\reals^{N \times N} \geq 0$,
there exists a real and positive eigenvalue $\lambda_1 =
\rho(\omat{A})$, called the \emph{Perron-Frobenius eigenvalue},
there exists a corresponding eigenvector $\ovec{v}$ to $\lambda_1$,
called the \emph{Perron-Frobenius eigenvector}, such that $\ovec{v}$
has positive components (it can always be picked to have Euclidean
norm equal to 1), and
the eigenvalue $\lambda_1$ is \textit{simple} in the sense that
$m_a(\lambda_1)=1$.


\section{Generalized-bias Opinion Models}
\label{sec.model}


This section presents a generalization of the DeGroot model to capture 
biases on opinion differences. It can account, among others, for dynamic influence/weights and inter-group
biases~\cite{hewstone2002} where agents favor one's own group over
other groups.
		
		
In social learning models, a community, society, or network is
typically represented as a directed weighted graph. The edges between
individuals (agents) specify the direction and strength of the
influence that one carries over the other.
			
\begin{definition}
	An ($N$-agent) \emph{influence graph} is a directed weighted graph
  $G=(A,E,I)$ with vertices $A = \left\{1,\ldots N\right\}$, edges
  $E \subseteq A \times A$, and weight function $I = A \times A \to
  \left[0,1\right]$ satisfying $I(i,j) = 0$ iff $(i,j) \notin E$.
\end{definition}

\noindent
The vertices $A$ represent the $N$ agents of a given community or
network.  Whenever $(i,j)\in E$, it means that agent $i$ influences
agent $j$. The value $I(i,j)$, for simplicity written $I_{i,j}$,
denotes the strength of the influence: $0$ means no influence and a
higher value means stronger influence.
The expression $A_{i}$ is used to denote the set of
(inbound-)neighbors of agent $i$, i.e., the set of agents $j$ with
$I_{j,i} > 0$. Recall that a graph is \emph{strongly connected} if
there is at least one path between any two vertices. In a strongly
connected influence graph, all the agents influence --directly or
indirectly-- one another.
			

The evolution of agents' opinions, about some underlying statement or
proposition, is modeled similar to the DeGroot-like models
in~\cite{degroot1974,Alvim2023,alvim-multiagentopinion-forte2024}.
			
An \emph{opinion state} (or \emph{belief state}) of the $N$ agents is
represented as a vector in $ \left[0,1\right]^{N}$. For any opinion
state $\ovec{x}= \oalgst{x_1,\ldots,x_N}$ and $1\leq i \leq N$, the
expression $\ovec{x}_{i}=x_i$ denotes the opinion of agent $i$.
If $\ovec{x}_{i}=0$, then $i$ completely disagrees with the
proposition; if $\ovec{x}_{i}=1$, then $i$ completely agrees with it.
The higher the value of $\ovec{x}_{i}$, the stronger the agreement.
			
\begin{definition}
  An \emph{opinion model} is a tuple
  $M=\left(G,\ovec{x}^{0},\mu_{G}\right)$, where G is an N-agent
  influence graph, $\ovec{x}^{0}$ is the \emph{initial state of
  opinion}, and $\mu_{G}: \left[0,1\right]^{N} \to
  \left[0,1\right]^{N}$ is a state-transition function, called
  \emph{update function}. For every $t \in \nats$, $\ovec{x}^{t + 1} =
  \mu_{G}(\ovec{x}^{t})$ is the state of opinion at time $t + 1$ in
  $M$.
\end{definition}

An update function can be used to express any deterministic and
discrete transition from one opinion state to the next, possibly
taking into account the influence graph. Intuitively, these update
functions specify the reaction of an agent to the opinion
disagreements with each one of its influencers.

Consensus is a central property in social learning (e.g., the
inability to reach consensus is a sign of a polarized society).

\begin{definition}
  Let $\left(G,\ovec{x}^0,\mu_{G}\right)$ be an opinion model with $G
  = \left(A,E,I\right)$.  A subset of agents $B \subseteq A$ is said
  to \emph{converge} to an opinion value $v$ iff for every $i \in B$,
  $\lim_{t \to \infty} \ovec{x}^{t}_{i} = v$. The subset $B$ is said
  to \emph{converge to consensus} iff $B$ converges to some opinion
  value $v$.
\end{definition}

\subsection{DeGroot Update}
\label{sec.model.degroot.update}

The \emph{normalized influence}
$\overline{I_{j,i}} = \frac{I_{j,i}}{\sum_{k \in A_{i}} I_{k,i}}$
is used to model the opinion update of agent $i$ as a function of its
neighbors' opinions~\cite{alvim-multiagentopinion-forte2024}.
Then, the standard DeGroot model~\cite{degroot1974} is obtained by the
update function $\mu_G(\ovec{x})_i  =  \sum_{j \in A_i}\overline{I_{j,i}} \ovec{x}_{j}$ which can be equivalently formulated as in Equation~\ref{eq.model.update}:
\begin{align}\label{eq.model.update}
  \mu_G(\ovec{x})_i & =  \ovec{x}_{i} + \sum_{j \in A_i}\overline{I_{j,i}} (\ovec{x}_{j} - \ovec{x}_{i}). 
\end{align}

\noindent
If $j$ influences $i$, then $i$'s opinion would tend to move closer to
$j$'s. The \emph{disagreement term} $(\ovec{x}_{j} - \ovec{x}_{i}) \in
[-1,1]$ in Equation~\ref{eq.model.update} realizes this intuition.
If $(\ovec{x}_{j} - \ovec{x}_{i})$ is a negative term in the sum, the
disagreement can be thought of as contributing with a magnitude of
$|\ovec{x}_{j} - \ovec{x}_{i}|$ (multiplied by $\overline{I_{j,i}}$)
to \emph{decreasing} $i$'s belief in the underlying
proposition.
Similarly, if $(\ovec{x}_{j} - \ovec{x}_{i})$ is positive, the
disagreement contributes with the same magnitude but to
\emph{increasing $i$'s} belief.

It is known that in standard DeGroot, two conditions on the influence
graph are enough for the agents converge to consensus.
First, the all the agents are required to influence, directly or
indirectly, each other.
Second, each agent has to have some self-influence.

\begin{theorem}[\cite{degroot1974}]\label{consensus-degroot}
  Let $\left(G,\ovec{x}^0,\mu_{G}\right)$ be an opinion model with $G
  = \left(A,E,I\right)$ and $\mu_G$ defined by
  Eq. \ref{eq.model.update}. If $G$ is (1) strongly connected and (2)
  for each $i\in A$, $\overline{I_{j,i}}<1$ for some $j\in A_i$, then
  $A$ converges to consensus.
\end{theorem}

\subsection{Updates with Bias Factors}
\label{sec.model.biasupdate}

A broad class of update functions, generalizing the original one in
DeGroot, is introduced in this section.
More precisely, updates for an agent $i$ can \emph{weight}
disagreements with each one of its neighbors $j$ using functions
$\alpha_{ij}$ from opinion states to $[0,1]$ 
referred to as  \emph{(generalized) bias factors}. 
The same opinion difference can then be weighted differently by
bias factors, depending of opinions of agents $i$ and $j$. Thus, intuitively $\alpha_{ij}$ may also be seen  as \emph{changing} the influence of $j$ over $i$ depending on their opinions. 
 
As an example, assume that $0.5$ represents a neutral opinion about an
underlying proposition. If $\ovec{x}_i=0.4$ and $\ovec{x}_j=0.1$, the
agents can be viewed as somewhat disagreeing and strongly disagreeing, respectively, with the underlying
proposition. If $\ovec{x}_i=0.7$ and $\ovec{x}_j=0.4$, agent $i$ seems
to agree somewhat with the proposition, while $j$ does not. Alas, in
both cases $(\ovec{x}_{j} - \ovec{x}_{i})=0.3$. It could be desirable
to have $\alpha_{i,j}(\ovec{x})$ assigning a higher weight in the
first situation because these opinions come from a group that
disagrees with the proposition. (This cannot be achieved with the influence graph alone since the opinion of agents may change while their influence in an opinion model does not.) As a matter of fact, this is an
instance of in-group bias.

\begin{definition}
  \label{def.model.genbias}
  Let $\left(G,\ovec{x}^{0},\mu_{G}\right)$ be an opinion model with $G =
  \left(A,E,I\right)$. The function $\mu_{G}: \left[0,1\right] \to
  \left[0,1\right]$ is an \emph{generalized bias update} iff, for each
  $i \in A$, it is defined by:
  \begin{align}
    \label{eq.model.scalarmap}
    \mu_{G}(\ovec{x})_{i} & = \ovec{x}_{i} + \sum_{j \in A_{i}} \overline{I_{j,i}}\alpha_{i,j}(\ovec{x})(\ovec{x}_{j} - \ovec{x}_{i}),
	\end{align}
	with each bias factor $\alpha_{i,j}: \left[0,1\right]^{N} \to \left[0,1\right]$
  satisfying:
  \begin{enumerate}
	\item $\alpha_{i,j}$ is a continuous function,
  \item if $\ovec{x}_{i} \neq \ovec{x}_{j}$, then
    $\alpha_{i,j}(\ovec{x}) > 0$, and
  \item for each $i \in A$, there exists $j \in A_{i}$, such that
    $\alpha_{i,j}(\ovec{x}) < 1$.
	\end{enumerate}
\end{definition}

\noindent
The first condition on the bias factor $\alpha_{i,j}$ guarantees that
small changes in the state of opinion lead to small changes in the
factor result.
Without the second condition, the bias factor would cancel the
influence of the contact $j$, which may break the strong connectivity
of the graph.
The third and last condition reproduces the second condition in the
Theorem~\ref{consensus-degroot} in the original DeGroot model.

Note that the update function in Equation~\ref{eq.model.scalarmap} can
be expressed equivalently as:
\begin{align}
		\mu_{G}(\ovec{x})_{i} & = \left(1 - \sum_{j \in A_{i}} \overline{I_{j,i}}\alpha_{i,j}(\ovec{x})\right)\ovec{x}_i + \sum_{j \in A_{i}} \overline{I_{j,i}}\alpha_{i,j}(\ovec{x})\ovec{x}_j, 
\end{align}
				
Furthemore, Equation~\ref{eq.model.scalarmap} is a \textit{convex
  combination} of $\left[0,1\right]$ elements and thus
$\mu_{G}(\ovec{x})_i \in \left[0,1\right]$ because $\left[0,1\right]$
is a convex set.
Consequently, the generalized-bias update function can be expressed in
vectorial form as:
\begin{align}
  \label{eq.model.vectorialmap}
	\mu_{G}(\ovec{x}) & = \omat{A}(\ovec{x})\ovec{x},
\end{align}
\noindent
where $\omat{A}: \left[0,1\right]^{N} \to \reals^{N \times N}$ is a square matrix with entries 
\begin{align}
  \label{eq.model.vectorialmapfam}
	a_{i,j}(\ovec{x}) =   \left\{ \begin{array}{lcc}
													\vspace{0.5cm}
													1 - \sum_{k \in A_{i}} \overline{I_{k,i}}\alpha_{i,k}(\ovec{x}) & \quad \mbox{, if} & i = j\\
													\overline{I_{j,i}}\alpha_{i,j}(\ovec{x}) &  \quad \mbox{, if} & i \neq j.
												\end{array}
											\right.
\end{align}

This section is concluded by introducing the notion of generalized
bias opinion models.

\begin{definition}
  \label{def.model.genbiasmodel}
  Let $M=\left(G,\ovec{x}^0,\mu_{G}\right)$ be an opinion model with
  $G = \left(A,E,I\right)$. Then, $M$ is an \emph{generalized-bias
  (opinion) model} iff $\mu_{G}$ can be expressed as in
  Equation~\ref{eq.model.vectorialmap}.
\end{definition}

\section{Consensus and Expressiveness}
\label{sec.consensus}

This section presents a consensus result for generalized-bias opinion
models (Definition~\ref{def.model.genbiasmodel}).
It is the main theoretical result of the paper and extends the
DeGroot's consensus result (see Theorem~\ref{consensus-degroot}).
This section also shows how generalized-bias models capture the
opinion specification and dynamics of other models, including
state-of-the-art~\cite{Alvim2023,alvim-multiagentopinion-forte2024,degroot1974}.
It highlights dynamics that generalized-bias models can handle, but
the other models cannot.

			
\begin{lemma}
  \label{lem.consensus.right}
  Let $\left(G,\ovec{x}^0,\mu_{G}\right)$ be an generalized-bias
  model, with $G = \left(A,E,I\right)$ and $\mu_{G}(\ovec{x}) =
  \omat{A}(\ovec{x})\ovec{x}$. The matrix $\omat{A}(\ovec{x})$ is
  (right-)stochastic for all $\ovec{x} \in \left[0,1\right]^{N}$.
			\end{lemma} 

Recall that a matrix $\omat{A} = \oalgst{a_{ij}}\in \reals^{N\times
  N}$ is (right-)stochastic if it entries satisfy $a_{ij} \geq 0$ and
$\sum^{N}_{j = 1} a_{ij} = 1$. This guarantees that \emph{consensual} state
of opinion vectors (i.e., vectors in which all agents have the same
opinion) in generalized-bias models are \emph{fixed points} of their
update function. Intuitively, this means that once the agents are in
consensus, they will remain in consensus.

\begin{lemma}
  \label{lem.consensus.primmat}
  Let $\left(G,\ovec{x}^0,\mu_{G}\right)$ be a generalized-bias
  model, with $G = \left(A,E,I\right)$ and $\mu_{G}(\ovec{x}) =
  \omat{A}(\ovec{x})\ovec{x}$. If $G$ is a strongly connected graph,
  then for every $t>0$, $\omat{A}(\ovec{x}^t)$ is primitive.
\end{lemma} 

Genelarized-bias models have  \emph{unique} fixed points.
The primitivity of $\omat{A}(\ovec{x}_{t})$, for $t > 0$, indicates
that they are consensual vectors.
This is a key observation for proving that generalized-bias models
converge to consensus, as it ensures that for each state opinion
vector $\ovec{x}^{t}$, its update $\ovec{x}^{t + 1}$ is closer to
consensus than $\ovec{x}^{t}$.
			
\begin{theorem}[Consensus for Generalized-Bias Models]
  \label{thm.consensus.generalized}
  Let $\left(G,\ovec{x}^0,\mu_{G}\right)$ be a generalized-bias model,
  with $G = \left(A,E,I\right)$ and $\mu_{G}(\ovec{x}) =
  \omat{A}(\ovec{x})\ovec{x}$.  If $G$ is a strongly connected graph,
  then $A$ converges to consensus.
\end{theorem}

\noindent
 The connectivity constraint on $G$ ensures that each pair of agents
$(i,j)$ is connected, directly or indirectly. Hence, the opinion of
$i$ is in a certain way influenced by the opinion of $j$ and
viceversa.
The non-zero condition of the the bias factors ensures that no matter
the difference of opinions between two agents, if there is a
connection between them, this influence, and hence their disagreement,
will not be ignored but it would be dynamically weighted.
Intuitively, the generalized-bias models guarantee that in an
environment where all users can freely express their opinions and be
heard, they will eventually come to an agreement even if there are
certain cognitive biases that regulate the influences between them.

The proofs of the above lemmas and Theorem~\ref{thm.consensus.generalized} can be found in  Appendix~\ref{apd.proofs}.



\paragraph{Expressiveness of Generalized-Bias Models. } The work in~\cite{alvim-multiagentopinion-forte2024} introduces \emph{disagreement-bias
(opinion) models}. They are extensions of DeGroot's of the form
$\left(G,\ovec{x}^0,\mu_{G}\right)$ with
\begin{align}\label{sec.expr.disagreement}
  \mu_{G}(\ovec{x})_i & = \ovec{x}_i + \sum_{j \in A_{i}} \overline{I_{j,i}}\beta_{i,j}(\ovec{x}_j - \ovec{x}_i),
\end{align}	
\noindent
where each $\beta_{i,j}: \left[-1,1\right] \to \left[-1,1\right]$ is a function referred to as \emph{disagreement
bias}. Intuitively, $\beta_{i,j}(\ovec{x}_j - \ovec{x}_i)$ is a factor
that depends only on the difference of opinion between $j$ and $i$.

It is shown that the agents in a disagreement-bias model converge to consensus if its influence
graph  is strongly connected and   each $\beta_{i,j}$ function is  continuous and in
the region $\mbox{\textbf{R}} = \left\{(x,y) \in \left[-1,1\right]^{2}
: xy > 0 \mbox{ and } \abs{y} < \abs{x} \right\} \cup
\left\{(0,0)\right\}$. 
The models with all their bias functions in $\textbf{R}$ can capture a
wide range of biases, such as variations of confirmation bias as well
as the dynamics of receptive and resistant agents (i.e., agents
willing to change their opinion towards their contacts but with some
skepticism)~\cite{Aronson10}.

\begin{theorem}[\cite{alvim-multiagentopinion-forte2024}]
  \label{thm.expr.forte}
  Let $\left(G,\ovec{x}^0,\mu_{G}\right)$ with $G =
  \left(A,E,I\right)$ be a disagreement-bias model whose disagreement
  bias functions are continuous and in $\textbf{R}$. If $G$ is strongly connected,  $A$
  converges to consensus.
\end{theorem}

Theorem~\ref{thm.expr.forte} is actually a consequence of
Theorem~\ref{thm.consensus.generalized} from this paper.
It can be shown that the update functions of disagreement-bias models,
whose continuous bias functions are in $\omat{R}$, can be expressed as update
functions of generalized-bias models. It can also be shown that there
are updates functions from generalized-bias models that cannot be
expressed as the update functions of a disagreement-bias model. More precisely,

\begin{theorem}[Expressiveness]
  \label{thm.expr.strictincl}
  (1) Let $(G,\ovec{x}^0,\nu_{G})$ be a disagreement-bias model whose
  disagreement bias functions are continuous and in $\omat{R}$. Then,
  there exists a generalized-bias model $(G,\ovec{x}^0,\mu_{G})$, such
  that $\mu_{G}=\nu_{G}$.  (2) There exists a generalized-bias model
  $(G,\ovec{x}^0,\mu_{G})$ such that  every disagreement-bias
  model $(G,\ovec{x}^0,\nu_{G})$  satisfies $\mu_{G}\neq \nu_{G}$.
\end{theorem}

\paragraph{Proof.} Part (1). Given any 
disagreement-bias model $(G,\ovec{x}^0,\nu_{G})$ whose
  disagreement biases are continuous and in $\omat{R}$, define the
generalized-bias model $(G,\ovec{x}^0,\mu_{G})$ where:
\begin{align}
    \alpha_{i,j}(\ovec{x}) &= \left\{\begin{array}{lcc} \vspace{0.5cm}
    \frac{\beta_{i,j}(\ovec{x}_j - \ovec{x}_i)}{\ovec{x}_j -
      \ovec{x}_i} & \mbox{\quad, if} & \ovec{x}_j \neq \ovec{x}_i,\\ 0 &
    \mbox{\quad, if} & \ovec{x}_j = \ovec{x}_i. \end{array} \right.
\end{align}
Here, $\beta_{i,j}$ is the corresponding disagreement bias in
$(G,\ovec{x}^0,\nu_{G})$. It can be verified that $0 <
\alpha_{i,j}(\ovec{x}) < 1$. Also, $\alpha_{i,j}$ is continuous
because $\beta_{i,j}$ is continuous. Thus, $\alpha_{i,j}$ satisfies
the conditions for the bias factors in
Definition~\ref{def.model.genbias}; it can be verified that
$\mu_G=\nu_G$.

For Part (2), consider the
generalized-bias model $(G,\ovec{x}^0,\mu_{G})$, with $G = (A,E,I)$,
$A = \left\{1,2\right\}$, $E = \left\{(1,2),(2,1)\right\}$, and
$I_{12} = I_{21} = 1.0$, and:
\begin{align}
  \mu_{G}(\ovec{x})_{1} &= \ovec{x}_1 + \alpha_{12}(\ovec{x}_{1},\ovec{x}_{2})(\ovec{x}_{2} - \ovec{x}_{1}), &
  \mu_{G}(\ovec{x})_{2} &= \ovec{x}_2 + \alpha_{21}(\ovec{x}_{1},\ovec{x}_{2})(\ovec{x}_{2} - \ovec{x}_{1}).
\end{align}
\noindent Suppose that $\alpha_{12}(\ovec{x}_{1},\ovec{x}_{2}) =
\alpha_{21}(\ovec{x}_{1},\ovec{x}_{2}) =
\alpha(\ovec{x}_{1},\ovec{x}_{2})$, where
\begin{align}
  \label{eq.expr.exttpos}
    \alpha(\ovec{x}) = \alpha(\ovec{x}_1,\ovec{x}_2) =
    2\left(\ovec{x}_1 - \ovec{x}^2_1 + 0.25\right)\left(1 -
    0.9\abs{\ovec{x}_1 - \ovec{x}_2}\right).
\end{align}
\noindent
Towards of contradiction, suppose that there exists a
disagreement-bias model $(G,\ovec{x}^0,\nu_{G})$ such that
$\nu_{G}=\mu_{G}$. Then, the update function $\nu_{G}$ must have the
form
\begin{align}
  \nu_{G}(\ovec{x})_{1} &= \ovec{x}_1 + \beta_{12}(\ovec{x}_{2} - \ovec{x}_{1}) & \nu_{G}(\ovec{x})_{2} &= \ovec{x}_2 + \beta_{21}(\ovec{x}_{1} - \ovec{x}_{2}),
\end{align}
with $\beta_{12}(\ovec{x}_{1} - \ovec{x}_{2})=
\alpha_{12}(\ovec{x}_{1},\ovec{x}_{2})(\ovec{x}_{2} - \ovec{x}_{1})$
and $\beta_{21}(\ovec{x}_{2} - \ovec{x}_{1}) =
\alpha_{21}(\ovec{x}_{1},\ovec{x}_{2})(\ovec{x}_{2} -
\ovec{x}_{1})$. However, for the opinion states $\ovec{y} =
\left[0,0.2\right]$ and $\ovec{z} = \left[0.5,0.7\right]$, it follows
that $-0.2\alpha_{12}(0,0.2) = \beta_{21}(-0.2) =
-0.2\alpha_{12}(0.5,0.7)$. This is a contradiction because
$\alpha_{12}(0,0.2) = 0.41$ and $\alpha_{12}(0.5,0.7) = 0.82$.\qed

\section{Intergroup Bias: A Case Study}
\label{sec.inter}  

Intergroup bias is a well-documented phenomenon where individuals tend
to evaluate members of their own group more favorably than members of
other groups~\cite{hewstone2002}. This section will illustrate opinion dynamics under this bias as
a particular case of a generalized-bias model.

Consider a scenario involving ideologically distinct groups that
traditionally clash, such as \emph{progressives} (left-wingers),
\emph{moderates} (centrists), and \emph{conservatives}
(right-wingers).
To illustrate how these groups might be divided by their opinions on a
controversial proposition, the statement ``free markets promote
prosperity'' is used.
Division is represented by splitting the interval $\left[0,1\right]$
into disjoint sub-intervals, corresponding to each group's agreement
with the proposition.  Define $I_{1} = \left[0,0.45\right)$, $I_{2} = \left[0.45,0.55\right]$  and $I_{3} = \left(0.55,1\right]$ 
for the
  progressive, moderate and conservative opinion groups.
  
\paragraph{Assessment Function.}
To capture inter-group bias, the bias factor $\alpha_{ij}$ is defined
as an \textit{assessment} function $a(x,y)$ assigning a weight
depending on the groups $x$ and $y$ belong to.
Let $w_{i}(x)$ be the characteristic function of the set $I_i$ (i.e.,
it is 1 if $x\in I_i$ and otherwise 0) and
$u_{ij}(x,y)=w_{i}(x)w_{j}(y)$ (i.e., it is 1 if $x \in I_i$ and $y
\in I_j$, and otherwise 0).

Characteristic functions are used so that the assessment function,
$a(x, y)$, first detects the group of the first agent and then detects
the group of the second agent. This will help in calculating the value
that the first agent assigns to the opinion of the second.
For detecting the group of the evaluating agent, the $w_{i}(x)$
function is used. To calculate the assessment that this agent gives to
the second agent, the $u_{ij}(x,y)$ function is used.
It then proceeds by defining the functions that calculate the
assessments that the first agent has of the second agent depending on
their groups: $a_{1}(x, y)$ when $x$ is progressive, $a_{2}(x, y)$
when $x$ is moderate, and $a_{3}(x, y)$ when $x$ is conservative.
These functions are combined by multiplying each one by the function
that detects the group of $x$.

First, start with $x$ in the progressives group ($I_1$). The group of
$y$ is identified and the corresponding value is defined according to
function $a_1$:
\begin{align}
  \label{eq.inter.assesment1}
    a_{1}(x,y) & = u_{11}(x,y) + u_{12}(x,y)\varphi(y) + 0.5u_{13}(x,y),
\end{align}


\noindent
with $\varphi(y) = 1 - 5(y - 0.45)$.
If $y$ is a progressive opinion as well, then $u_{11}(x,y) = 1$ and
$u_{12}(x,y) = 0=u_{13}(x,y)$; thus the weight is $1$.
If $y$ is a moderate opinion, the weight is given by $\varphi(y)$; a
decreasing linear function stating that when the distance grows, the
weight decreases. Notice that $\varphi(0.45)=1$ and
$\varphi(0.55)=0.5$.
Finally, if $y$ is a conservative opinion the weight is equal to
$0.5$. Clearly, $a_1(x,y)$ is continuous in $x \in I_1$ and $y \in
[0,1].$



Second is the case when $x$ is a moderate opinion ($I_2$), handled
with $a_2$:
\begin{align}
  \label{eq.inter.assesment2}
  a_{2}(x,y) & =\varphi(x)u_{21}(x,y) + \kappa(x,y)u_{22}(x,y) + \psi(x)u_{23}(x,y),
\end{align}
with $\psi(x) = 0.5 + 5\left(x - 0.45\right)$ and $\kappa(x,y) = 100xy
- 50x - 50y + 25.75$. It can be checked that $\kappa(0.45,y) =
\varphi(y)$ and $\kappa(0.55,y) = \psi(y)$; thus, $a_2(x,y)$ is
continuous function in $x\in I_2$ and $y \in [0,1]$.

Third, the case when $x$ is a conservative opinion ($I_3$). The
assessment is given by function $a_3$:
\begin{align}
  \label{eq.inter.assesment3}
  a_{3}(x,y) & = 0.5u_{31}(x,y) + \psi(y)u_{32}(x,y) + u_{33}(x,y),
\end{align}
which is continuous in $x\in I_3$ and $y\in [0,1]$.

\begin{figure}
  \centering
  \includegraphics[width=1\linewidth]{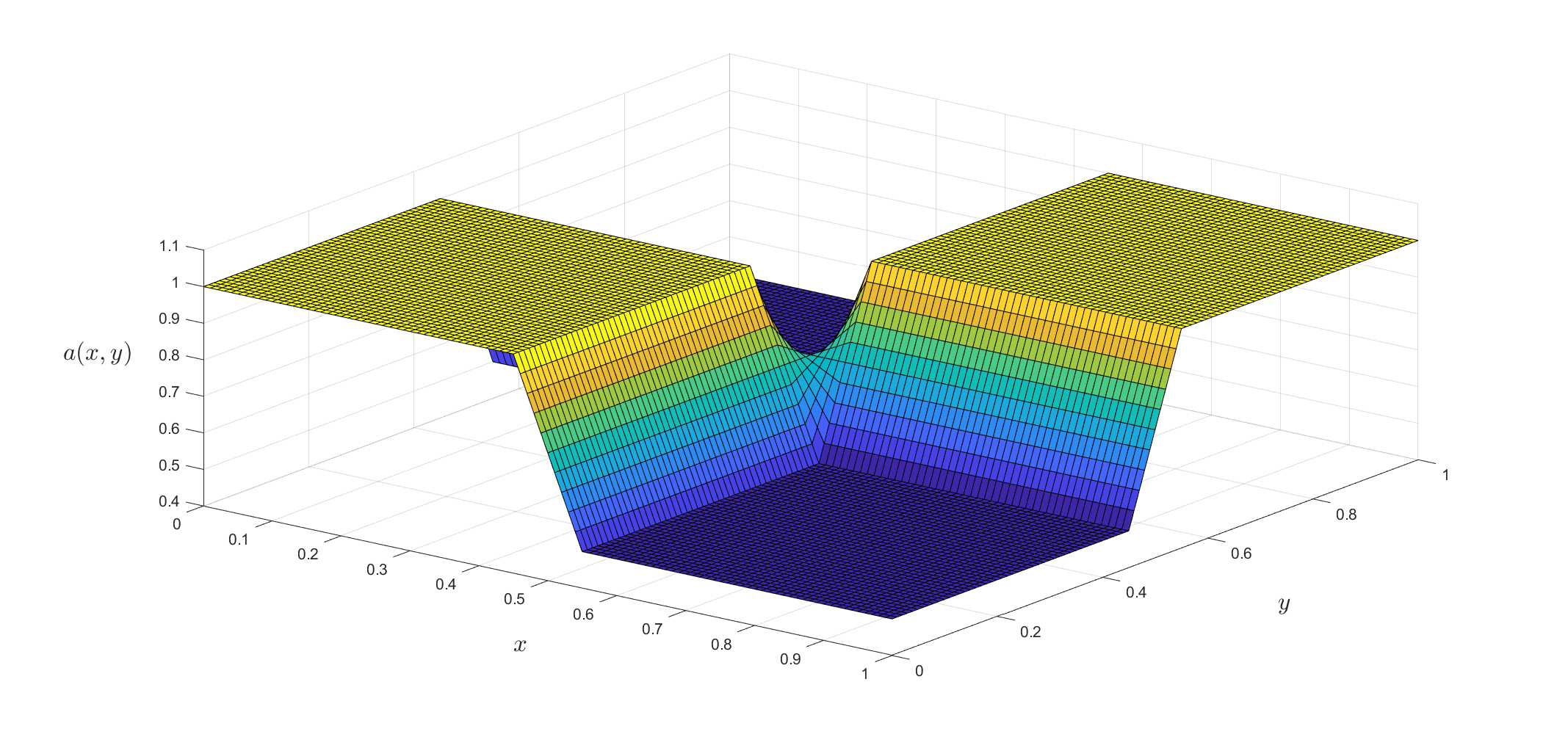}
  \caption{Surface of function $a(x,y)$ defined in Eq.~\ref{eq.inter.assesment}.}
  \label{fig.inter.assesment}
\end{figure}

With this setup, the continuous assessment function $a$ (see
Figure~\ref{fig.inter.assesment}):
\begin{align}
  \label{eq.inter.assesment}
  a(x,y) & = w_{1}(x)a_{1}(x,y) + w_{2}(x)a_{2}(x,y) + w_{3}(x)a_{3}(x,y). 
\end{align}

Note that function $a(x,y)$ is continuous, positive for all $(x,y) \in \left[0,1\right]^{2}$, and if each agent in the graph is connected to a member of another group as in the example below, then the assessment function is indeed a bias factor
(Definition \ref{def.model.genbias}).

The update function for inter-group bias is finally defined as:

\begin{align}
  \label{eq.inter.update}
  \mu_{G}(\ovec{x}) = \ovec{x}_i + \sum_{j \in A_{i}} \overline{I_{j,i}}\alpha_{ij}(\ovec{x}_i,\ovec{x}_j)(\ovec{x}_i - \ovec{x}_j),
\end{align}
where $\alpha_{ij}(\ovec{x}_i,\ovec{x}_j) = a(\ovec{x}_i,\ovec{x}_j)$. 

The following example illustrates evolution of opinion under this form of inter-group bias. 

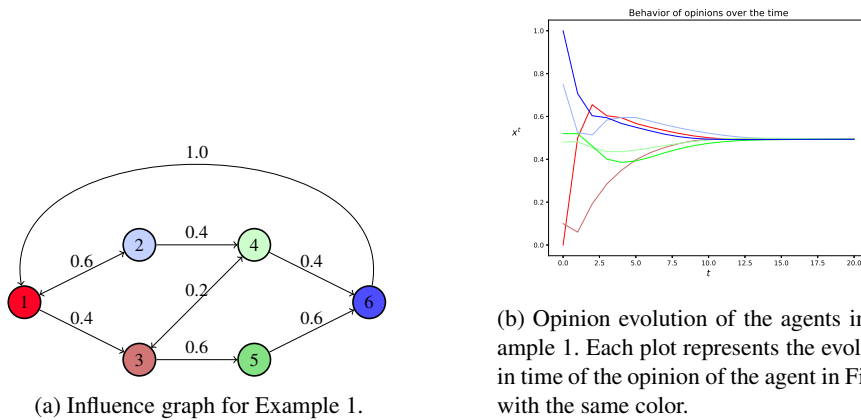
\begin{figure}[htbp]
        
        
  \begin{subfigure}[b]{0.45\textwidth}
    \centering
    \resizebox{\columnwidth}{!}{%
      \begin{tikzpicture}
        \tikzstyle{every state}=[fill opacity=1.0,text opacity=1,thick,minimum size=12pt]
        
        \definecolor{darkred1}{rgb}{1.0,0.01,0.15}
        \definecolor{blue3}{rgb}{0.6, 0.7, 1.0}
        \definecolor{red2}{rgb}{0.8, 0.4, 0.4}
        \definecolor{green1}{rgb}{0.2, 0.8, 0.2}
        \definecolor{darkblue1}{rgb}{0.0, 0.0, 1.0}
        
        \node[state, fill=darkred1!100] (0) at (0,0) [] {1};
        \node[state, fill=blue3!60] (1) at (2,1) [] {2};
        \node[state, fill=red2!90] (2) at (2,-1) [] {3};
        \node[state, fill=green3!50] (3) at (4,1) [] {4};
        \node[state, fill=green1!60] (4) at (4,-1) [] {5};
        \node[state, fill=darkblue1!70] (5) at (6,0) [] {6};
        
        \draw
        (0) edge[<->] node[pos=0.5, above] {0.6} (1)
        (1) edge[->]  node[pos=0.5, above] {0.4} (3)
        (3) edge[->]  node[pos=0.5, above] {0.4} (5)
        (0) edge[->]  node[pos=0.5, above] {0.4} (2)
        (2) edge[->]  node[pos=0.5, above] {0.6} (4)
        (4) edge[->]  node[pos=0.5, above] {0.6} (5)
        (3) edge[<->] node[pos=0.5, above] {0.2} (2)
        (5) edge[->, looseness=1.2, bend right=100] node[pos=0.5, above] {1.0} (0);
      \end{tikzpicture}
    }
    \caption{Influence graph for Example~\ref{exa.inter.example}.}
    \label{fig.inter.graph}
  \end{subfigure}
  \hfill
  \begin{subfigure}[b]{0.45\textwidth}
    \centering
    \includegraphics[width=\linewidth]{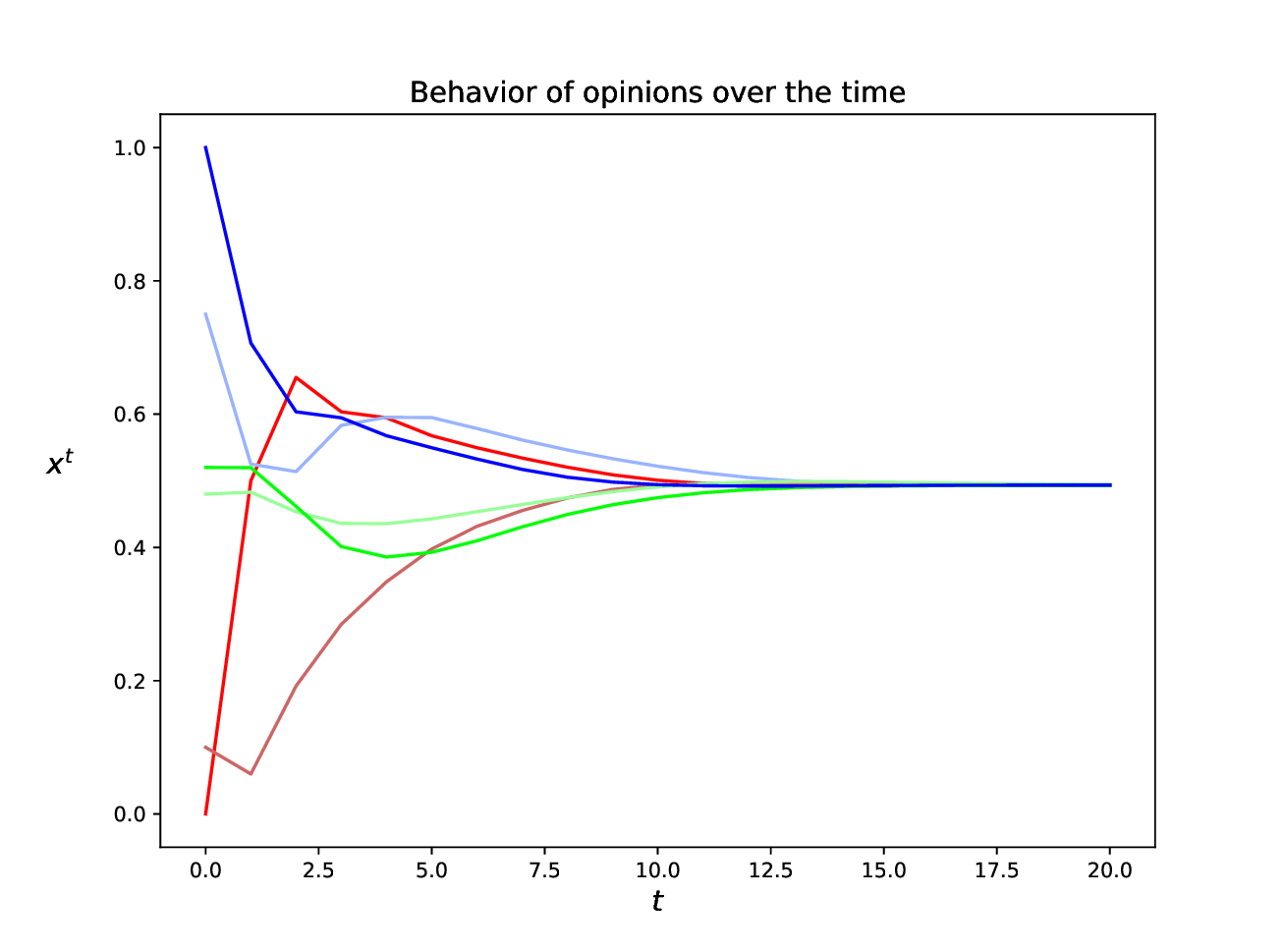}
    \caption{Opinion evolution of the agents in Example~\ref{exa.inter.example}. Each plot represents the evolution in time of the opinion of the agent in Fig.~\ref{fig.inter.graph} with the same color.}
    \label{fig.inter.time}
  \end{subfigure}
  \caption{Influence graph and opinion evolution.}
\end{figure}

\begin{example}
  \label{exa.inter.example}
  Let $\left(G,\ovec{x}^0,\mu_{G}\right)$ be a generalized-bias model
  for six agents where $G$ is the graph in Figure~\ref{fig.inter.graph}.
  The initial state of opinion is $\ovec{x}^0 =
  \oalgst{0.0,0.75,0.1,0.48,0.52,1.0}$ and $\mu_{G}$ is the
  inter-group bias update in Equation~\ref{eq.inter.update}.
  Initially, there are two agents that are progressives (agents in
  red), two moderates (agents in green), and two conservatives (agents
  in blue).
  Note that $G$ is strongly connected. Because of the influence graph
  and the initial state opinions, there are interactions between
  agents of different groups, as well as in the same group.
  Figure~\ref{fig.inter.time} illustrates the evolution of opinion of
  the agents converging to consensus (around a moderate opinion) as
  predicted by the consensus result in
  Theorem~\ref{thm.consensus.generalized}.
\end{example}

\section{Concluding Remarks}
\label{sec.concl}

This paper introduced the notion of generalized-bias opinion dynamic
model, a generalization of the classic DeGroot model, in which agents
may react differently to the same opinion difference depending on
whether the person holding the opposing view is from the same group.
These reactions were formalized as arbitrary functions --parametric to
the models-- that depend, not only on opinion difference, but also on
 the opinions of the individuals interacting.
This is a well-known bias in social psychology that cannot be modeled
by state-of-the-art DeGroot-based models, as shown in
Theorem \ref{thm.expr.strictincl}. Furthermore, a consensus result  was provided (Theorem \ref{thm.consensus.generalized}), extending the classic result of the DeGroot model \cite{degroot1974} for a wide range of social biases. Finally, an application example involving a particular case study of inter-group bias was used to illustrate the  expressiveness of the new model.

%

The relevance of biased reasoning in human interactions has been extensively studied in \cite{Psychology1}, \cite{Psychology2}, \cite{Psychology3}, and others. Previous research in social psychology has explored intergroup bias~\cite{gaertner1999,gaertner2014,pettigrew2008}, concluding that fostering dialogue between members of different groups reduces the disparity in their opinions. These findings align with the consensus results and the case study presented in the current work. The present model may bring further insights to social scientists about cognitive bias in opinion formation. 

There is a great deal of work on formal models for belief change in social networks; we focus on the work on biased belief updates, which is the focus of this paper. Some models have previously been proposed to generalize the DeGroot model and introduce bias. For instance, \cite{Generalize1}, \cite{Generalize2}, and \cite{Generalize3} analyze the effects of incorporating a bias factor for each agent to represent biased assimilation—how much of the external opinions the agent will take into consideration. The main difference between these models and the generalized-bias models in this paper is that in the previous models, biases are not represented as arbitrary continuous functions satisfying two conditions. Instead, biases are either modeled as an exponential factor that reduces the impact of neighbors' opinions or by dynamically changing the weights in the DeGroot model with a specific function. Thus, the new model provides greater flexibility for capturing a wider range of biases.

The generalized-bias opinion models introduced here assume synchronous communication among the agents in the network, meaning all agents update their opinions simultaneously. In the line of~\cite{olarte-unifiedopinion-wrla2024,aranda-fariness-concur2024,haar:hal-01256984}, agents
can also communicate asynchronously  or
via a hybrid blend where both types of communication can coexist. It
would be interesting to see if the results presented here can be
extended to these two settings.
Moreover, it would be worth pursuing the use of rewriting logic  to specify generalized-bias opinion
models and perform several forms of formal analysis, including
reachability analysis, model checking, and statistical model checking
of concrete instances and desired properties of these models. Finally, the model could be extended to include agents that can learn by exchanging beliefs and lies as done in  \cite{haar:hal-01256984}.

\bibliographystyle{abbrv}
\bibliography{biblio}

\appendix 

\newpage

\section{Matrices and Dynamical Systems}
\label{sec.prelimapp}

\subsection{Matrices}
\label{sec.prelimapp.stoc}

Square matrices in $\reals^{N \times N}$ can be seen as
representations of directed graphs with $N$ vertices and vice-versa.

\begin{definition}\label{def.prelim.graph}
  The graph $G_\omat{A}$ \emph{associated} to a matrix $\omat{A} =
  \oalgst{a_{ij}} \in \reals^{N \times N}$ is defined by $G_\omat{A} =
  (V_\omat{A}, E_\omat{A})$ with $N$ nodes $V_\omat{A} = \{1, \ldots,
  N\}$ and edges $E_\omat{A} = \{ (i, j) \mid a_{ij} = 1 \}$.
\end{definition}

Structural properties of (square) matrices may be studied from the
structure and connectivity in the associated graphs. Recall that a
graph is \emph{strongly connected} if it has exactly one strongly
connected component (i.e., there is a path between any pair of its
nodes).

\begin{theorem}\label{thm.prelim.redstcon}
  A matrix $\omat{A}$ is irreducible iff $G_\omat{A}$ is strongly
  connected.
\end{theorem}

Stochastic matrices are a special case of non-negative matrices. They
are also known as a probability matrices, transition matrices, or
Markov matrices, and are used to describe the transitions of a Markov
chain.

\begin{definition}\label{def.prelim.stochastic}
  A matrix $\omat{A} = \oalgst{a_{ij}}\in \reals^{N\times N}$ is
  called \emph{(right-)stochastic} iff the following to conditions
  hold on $\omat{A}$:
  \begin{enumerate}
  \item
    It is non-negative.
  \item
    The sum of each one of its rows is 1, i.e., for each $1 \leq i
    \leq N$, it satisfies: \[\sum_{j=1}^N a_{ij} = 1.\]
  \end{enumerate}
\end{definition}

It can be checked that any stochastic matrix $\omat{A}$ has an
eigenvalue equal to 1, which corresponds to its spectral radius. This
follows from the fact that the sum of each row of $\omat{A}$ is 1, and
hence there is an eigenvalue equal to 1 associated to a stationary
distribution vector (i.e., a probability vector) $\ovec{v}$ satisfying
$\omat{A}\ovec{v} = \ovec{v}$.

\subsection{Dynamical Systems}
\label{sec.prelim.ds}

A dynamical system is a formal concept used to describe a system that
evolves over time according to a specific set of rules.
%

\begin{definition}\label{sec.def.ds}
  A \emph{(discrete) dynamical system} is a pair $\scal = (S, \phi)$
  where $S$ is a set and $\phi = \{\phi_t\}_{t\in T}$ is a family of
  $T$-indexed functions $\ofunc{\phi_t}{S}{S}$, with $T$ a semigroup,
  satisfying for any $x \in S$ and $s,t \in T$:
  \begin{enumerate}
  \item
    $\phi_0(x) =  x$.
  \item
    $\phi_s(\phi_t(x)) = \phi_{s+t}(x)$.
  \end{enumerate}
  For convenience, the expression $\phi_t(x)$ is abbreviated $x_t$,
  for any $x\in S$ and $t\in T$.
\end{definition}

In a dynamical system $\scal = (S, \phi)$, the set $S$ represents all
possible states of the system and the iterative application of $\phi$
defines the evolution of the system.
The \emph{orbit} of a state $x \in S$ is the set $\{ x_t \mid t \in
T\}$ of all states reachable from $x$.
A set $\Lambda \subseteq S$ is called an \emph{invariant set} of
$\scal$ iff $\phi_t(x) \in \Lambda$ for any $x \in \Lambda$ and $t \in T$.
An invariant set is a subset of the state space that, once entered by
a trajectory of the system, it cannot be left. In other words, if the
system's state ever enters an invariant set, it will remain in that
set for all future time steps.
Since an invariant set is a subset of the state space that is closed
under transitions, it can include the orbits of many elements in the
state space.

For the purpose of this work, the set $S$ is assumed to be $\reals^N$
and the index set $T$ is the set of natural numbers $\nats$.

\begin{example}\label{exa.prelim.dynsys}
  Consider the dynamical system $\scal = (\reals^2, \phi)$, with $\phi
  = \{ f^n \}_{n \in \nats}$ and $\ofunc{f}{\reals^2}{\reals^2}$
  defined by:
  \begin{equation*}
    f(\ovec{x})
    =  \omat{A}\ovec{x}
    = \left[
      \begin{array}{lcc}
        0.5 & 0.5 \\
        0.2 & 0.8
      \end{array}
      \right]
    \ovec{x}.
  \end{equation*}
Note that $f$ is the application of the irreducible stochastic matrix
$\omat{A}$ on a given vector $\ovec{x}$. By the Perron-Frobenius
Theorem (see Section~\ref{sec.prelim}), the spectral radius
$\rho(\omat{A})=1$ has multiplicity 1. Therefore, there is exactly one
eigenvector $\ovec{v}$ (modulo scalar multiplication) satisfying
$\ovec{x} = f(\ovec{x})$. Consequently, the identity over $\reals^2$
is an invariant set for $\scal$.
\end{example}

An attractor in a dynamical system is an invariant set toward which
the system tends to evolve from a wide variety of initial
conditions. Attractors can play a crucial role in understanding the
long-term behavior of dynamical systems.

\begin{definition}\label{def.prelim.attractor}
  Let $\scal = (S, \phi)$ be a dynamical system. An \emph{attractor}
  is a set of states $U \subseteq S$ with the following properties:
  \begin{enumerate}
  \item
    $U$ is an invariant set.
  \item
    For any point $x \in U$, the distance from $x_n$ to $U$ tends to
    zero as $n \in \nats$ tends to infinity.
  \end{enumerate}
\end{definition}

Attractors can be fixed points, periodic orbits, limit cycles, or more
complex structures like strange attractors. They are invariant and
have a \emph{basin of attraction} that draws nearby trajectories,
making them key to understanding the system's long-term behavior.

An \emph{$\omega$-limit set} (often called an \emph{$\omega$-set}) is
a set of accumulation points that a trajectory approaches as time goes
to infinity. It also provides important information about the
long-term behavior of the system. Formally, for $\scal = (S, \phi)$ a
dynamical system, the \emph{$\omega$-limit set} of a point $x \in S$,
denoted $\omega(x)$, is defined as the set of all points $y \in S$,
such that there exists an (strictly) increasing sequence $\{n_k\}_{k
  \in \nats}$ satisfying $\lim_{k \to \infty} \phi_{n_k}(x) = y$.

\begin{example}
  Consider the dynamical system $\scal = ([0..1], \phi)$, with $\phi =
  \{ f^n \}_{n \in \nats}$ and $\ofunc{f}{\reals}{\reals}$ the
  \emph{logistic map} defined by $f(x) = \mu x(1 - x)$.

  \begin{figure}[htbp]
    \centering
    \includegraphics[scale = 0.6]{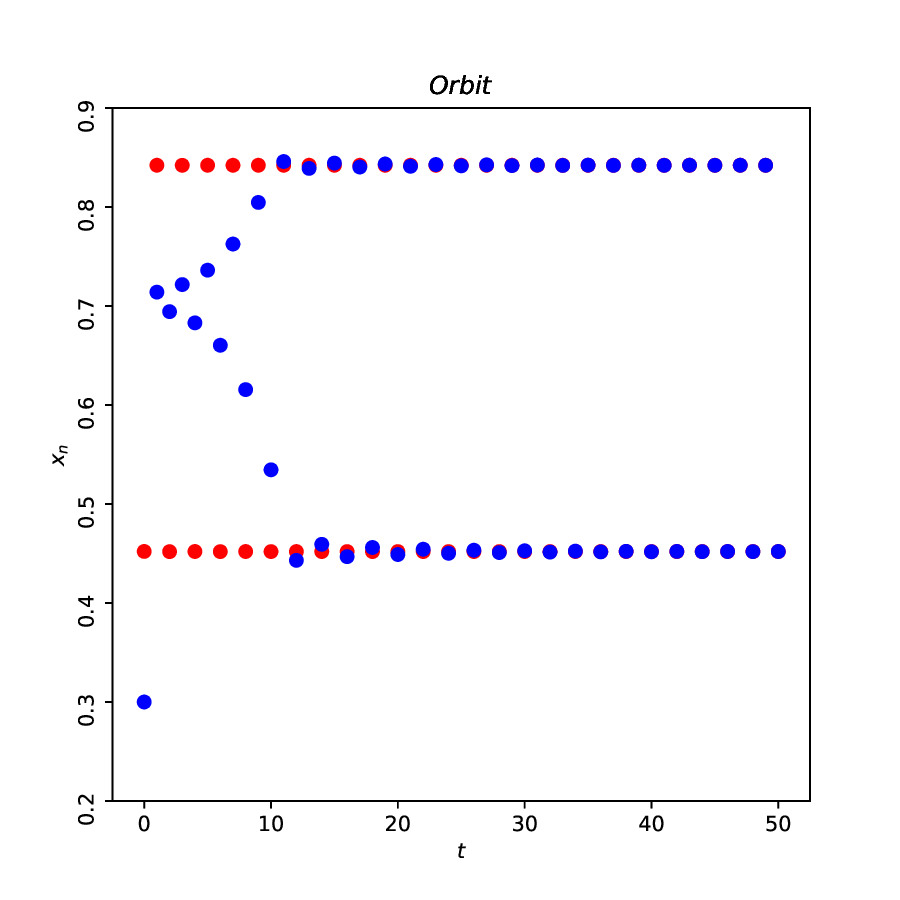}
    \caption{Orbit of point $x_{0}$ under map $L$, with $\mu = 3.4$ in
      blue. Set $\left\{0.4519,0.8421\right\}$ in red.}
    \label{fig.prelim.orbit}
  \end{figure}

  Figure~\ref{fig.prelim.orbit} depicts the orbit of $x_0$ in $\scal$
  when setting $\mu = 3.4$ and $x_{0} = 0.3$.
  It converges to the set $U = \left\{0.4519,0.8421\right\}$ (the red
  dots). The even indexes converge to $0.4519$ and the odd indexes
  converge to $0.8421$. That is, $\omega(x_{0}) = U$.
\end{example}

\newpage

\section{Proofs}
\label{apd.proofs}

In this appendix, we present the proofs of our theoretical results for consensus.

\subsection{Lemma \ref{lem.consensus.right}}

To prove this result, we have to prove that for all $\ovec{x} \in \left[0,1\right]^N$, the entries $a_{ij}(\ovec{x})$ of matrix $\omat{A}(\ovec{x})$, satisfy

\begin{itemize}
    \item $a_{ij}(\ovec{x}) \geq 0$, for all $i,j = 1,\ldots N$.
    \item $\sum^{N}_{j = 1} a_{ij}(\ovec{x}) = 1$, for all $i = 1,\cdots N$.
\end{itemize}

We begin the proof with the definition of these entries, given in equation \eqref{eq.model.vectorialmapfam}

\begin{align}
	a_{i,j}(\ovec{x}) =   \left\{ \begin{array}{lcc}
													\vspace{0.5cm}
													1 - \sum_{k \in A_{i}} \overline{I_{k,i}}\alpha_{i,k}(\ovec{x}) & \quad \mbox{, if} & i = j\\
													\overline{I_{j,i}}\alpha_{i,j}(\ovec{x}) &  \quad \mbox{, if} & i \neq j.
												\end{array}
											\right.
\end{align}

Note that $a_{i,i} = 1 - (a_{i,1} + \ldots a_{i,i - 1} + a_{i,i + 1} + \ldots a_{i,N})$, then, it is easy to see that $a_{i,1} + \ldots a_{i,i - 1} + a_{i,i} + a_{i,i + 1} + \ldots a_{i,N} = 1$. 

Now, note that $a_{i,j} \geq 0$, $i \neq j$, since $\overline{I}_{j,i} \geq 0$ and $\alpha_{i,j}(\ovec{x}) \geq 0$, for all $i,j = 1,\ldots N$.

Recall that $\sum_{j \in A_{i}} \overline{I}_{j,i} = 1$ and $\alpha_{i,j}(\ovec{x}) \leq 1$, furthermore, for each $i \in A$, there exist a neighbor $j \in A_{i}$ such that $\alpha_{i,j}(\ovec{x}) < 1$, then 

\begin{equation*}
    \sum^{N}_{k \in A_{i}} I_{k,i}\alpha_{i,k}(\ovec{x}) < \sum^{N}_{k \in A_{i}} I_{k,i} = 1,
\end{equation*}

with this, we can check that $a_{i,i} = 1 - \sum^{N}_{k \in A_{i}} I_{k,i}\alpha_{i,k}(\ovec{x}) > 0$.

Hence $\omat{A}(\ovec{x})$ is stochastic, for all $\ovec{x} \in \left[0,1\right]^N$.

\subsection{Lemma \ref{lem.consensus.primmat}}

To prove that matrix $\omat{A}(\ovec{x})$ is primitive, we have to check several conditions

\begin{itemize}
    \item $\omat{A}(\ovec{x})$ is irreducible for all $\ovec{x} \in \left[0,1\right]^N$.
    \item There exists $m > 0$, such that $\omat{A}^m(\ovec{x})$, for all $\ovec{x} \in \left[0,1\right]^N$.
\end{itemize}

By Theorem \ref{thm.prelim.redstcon}, prove that for all $\ovec{x} \in \left[0,1\right]^N$, $\omat{A}^m(\ovec{x})$ is irreducible, is equivalent to prove that the graph $G_{\omat{A}(\ovec{x})}$ associated at $\omat{A}(\ovec{x})$, is strongly connected.

By hypothesis, the graph $G$, is strongly connected, then for each pair of vertex $i,j$, there exist a path of edges $(i,v_{2})\ldots(v_{k},j)$ which connects these vertex.

It is easy to see that $I_{j,i} > 0$, implies $a_{i,j} > 0$, for $i \neq j$. Then, for each edge $(j,i)$ in $E$, there exists an edge between vertex $(i,j)$, in $G_{\omat{A}(\ovec{x})}$. From this, we can conclude that for each pair of vertex $(i,j)$ in $G_{\omat{A}(\ovec{x})}$, exists a path of edges which connects these vertex, hence $G_{\omat{A}(\ovec{x})}$ is strongly connected.

Above argument show that $\omat{A}(\ovec{x})$ is irreducible for all $\ovec{x} \in \left[0,1\right]^N$.

To prove the second condition, we shall use the Frobenius test of primitivity with $m = \abs{A}$. If we denote $a^{(m)}_{i,j}$ like the entry $i,j$ of matrix $\omat{A}^m$, then we must prove that $a^{(m)}_{i,j} > 0$, for all $i,j = 1,\ldots N$.

We shall prove this with the next two claims

\textbf{Claim 1:} If exist a path of length $l$ between $i$ and $j$, then $a^{(l)}_{i,j} > 0$.

\textbf{Proof:} We prove this claim by induction over $l$. 

The base case is $l = 1$, what is trivial, since the existence of an edge $(i,j)$ implies $a_{i,j} > 0$.

Inductively, assume that if exist a path of length $l$ between 2 nodes $i$, $k$, then $a^{(l)}_{i,k} > 0$. 

Consider a pair of vertex $i$, $j$ and a path of edges $(i,v_{1})(v_{1},v_{2})\ldots (v_{l - 1}.v_{l})(v_{l},j)$ of length $l + 1$. For inductively hypothesis $a^{(l)}_{i,v_{l}} > 0$, then

\begin{equation*}
    a^{(l + 1)}_{i,j} = \sum^{N}_{k = 1} a^{(l)}_{i,k}a_{k,j} \geq a^{(l)}_{i,v_{l}}a_{v_{l},j} > 0.
\end{equation*}

It is proved.

\textbf{Claim 2:} If $a^{(n)}_{i,j} > 0$, then $a^{(n + 1)}_{i,j} > 0$, for all $n \geq 1$.

\textbf{Proof:} Recall that for all $j = 1,\ldots N$, $a_{j,j} > 0$, then, if we have $a^{(n)}_{i,j}$, for some $n \geq 1$, we have

\begin{equation*}
    a^{(n + 1)}_{i,j} = \sum^{N}_{k = 1} a^{(n)}_{i,k}a_{k,j} \geq a^{(n)}_{i,j}a_{j,j} > 0.
\end{equation*}

It is proved.


Since for all $\ovec{x} \in \left[0,1\right]^N$, $G_{\omat{A}(\ovec{x})}$ is strongly connected, then for each pair of vertex $(i,j)$ exists a path of length $l$, where $l \leq \abs{A}$. By Claim 1, $a^{(l)}_{i,j} > 0$, and by Claim 2, $a^{(\abs{A})}_{i,j} > 0$. Then $\omat{A}^{\abs{A}}(\ovec{x}) > 0$, and by Frobenius test of primitivity, $\omat{A}(\ovec{x})$ is primitive.

\subsection{Theorem \ref{thm.consensus.generalized}}

The proof of this result requires several concepts from dynamical systems, as well as the lemmas stated above. Recall that an update function converges to consensus if there exists a $v \in \left[0,1\right]$ such that, for every agent $i \in A$, their opinion $\ovec{x}^t_i$ satisfies $\lim_{t} \ovec{x}^t_i = v$.

We equivalently prove that the sequence $\left\{\eta_{k}\right\} \subset \R$, defined by $\eta_{k} = \norm{\ovec{x}^k - \overline{\ovec{x}^k}}$, where $\ovec{x}^{k+1} = \mu_{G}(\ovec{x}^k)$, converges at $\eta_{k} \to 0$.

Proof is composed of the next way

\begin{itemize}
    \item Prove that the sequence $\eta_{k} = \norm{\ovec{x}^k - \overline{\ovec{x}}_k}$ converges at one value $L \geq 0$ when $k \to \infty$.
    \item Prove that $L = 0$.
\end{itemize}

Consider the line $l \subset \Rn$, defined by $l = \left\{r \ovec{1} : r \in \R \right\}$, where $\textbf{1}$ is a vector whose all entries are equal to 1. 

\textbf{Claim 1:} The point $\ovec{y} = \overline{\ovec{x}} \in l$, is the point who minimize the distance $\norm{\ovec{x} - \textbf{y}}$.

\textbf{Proof:} Proof of above claim is a geometric property, since $\overline{\ovec{x}}$ is the projection of $\ovec{x}$ over the line $l$.

\textbf{Claim 2:} The sequence $\eta_{k} = \norm{\ovec{x}^k - \overline{\ovec{x}^k}}$, where $\ovec{x}^{k+1} = \mu_{G}(\ovec{x}^k)$, converges at converges at one value $L \geq 0$.

\textbf{Proof:} To prove this claim, we have to prove that the sequence $\eta_{k}$ is monotonous and bounded. 

First, we use the Claim 1 and that vectors in $l$ are fixed points of map $F$

\begin{equation*}
    \begin{split}
        \norm{\ovec{x}^{k+1} - \overline{\ovec{x}^{k+1}}} &< \norm{\ovec{x}^{k+1} - \overline{\ovec{x}^k}},\\ 
        \norm{\ovec{x}^{k+1} - \overline{\ovec{x}^{k+1}}} &< \norm{\omat{A}(\ovec{x}^k)\ovec{x}^k - \omat{A}(\ovec{x}^k)\overline{\ovec{x}^k}},\\ 
        \norm{\ovec{x}^{k+1} - \overline{\ovec{x}^{k+1}}} &< \norm{\omat{A}(\ovec{x}^k)}\norm{\ovec{x}^k - \overline{\ovec{x}^k}},\\ 
    \end{split}
\end{equation*}

Vector $\ovec{x}^k - \overline{\ovec{x}^k}$ is in the space generated by the eigenvectors $\ovec{v}_{2},\ldots \ovec{v}_s$, $s \geq N$. This space is invariant for the operator $\omat{A}$, then the norm of this operator is $\abs{\lambda_{2}} < 1$ (because $\omat{A}$ is primitive), then

\begin{equation*}
    \begin{split}
        \norm{\ovec{x}^{k+1} - \overline{\ovec{x}^{k+1}}} &< \abs{\lambda_{2}}\norm{\ovec{x}^k - \overline{\ovec{x}^k}},\\ 
        \norm{\ovec{x}^{k+1} - \overline{\ovec{x}^{k+1}}} &< \norm{\ovec{x}^k - \overline{\ovec{x}^k}},\\ 
    \end{split}
\end{equation*}

With this, we have shown that the sequence is decreasing monotone. 

By properties of the norm operator $\norm{.}$, this is non negative, for this reason and the monotonic property, the sequence is bounded. For each $k \geq 1$, we have $0 \leq \eta_{k} \leq \eta_{0}$. By the real analysis, we know that this sequence converges at some value $L$, where $0 \leq L \leq \eta_{0}$.

\textbf{Claim 3:} The limit $L$ of the sequence $\eta_{k}$ is $L = 0$.

The first item to prove this claim, is that the update function $\mu_{G}$ defined in \eqref{eq.model.vectorialmap} is continuous. We know this because this map is composed by continuous functions.

For properties of continuous dynamical systems presented in Appendix \ref{sec.prelim.ds}, we have that for each initial condition $\ovec{x}^0$, the set $\omega_{F}(\ovec{x}^0)$ is an invariant set; which means that for all $\textbf{y} \in \omega_{F}(\ovec{x}^0)$, $F(\textbf{y}) \in \omega_{F}(\ovec{x}^0)$.

Now, let us proceed by contradiction. Suppose that $L > 0$, then $\norm{\ovec{x}^k - \overline{\ovec{x}^k}} \to L > 0$, when $k \to \infty$. This implies that $\omega_{\mu}(\ovec{x}^0)$ is a subset of the sphere of radius $L$ centered in a point $\ovec{p} \in \Rn$ of the line $l$, denoted by $S_{L}$.

Let $\ovec{y} \in \omega_{\mu}(\ovec{x}^0)$, by our above reasoning, we have that $\norm{\ovec{y} - \overline{\ovec{y}}} > \norm{\mu_{G}(\ovec{y}) - \overline{\mu_{G}(\ovec{y})}}$, but this contradicts the invariance of $\omega_{\mu_{G}}(\ovec{x}^0)$, because the distance of $F(\ovec{y})$ is less than $L$.

In conclusion, $L$ must be equal to zero.

\end{document}